\documentclass[useAMS,usenatbib]{mn2e}

\usepackage{psfig}

\usepackage{txfonts}

\title[Thermodynamics of molecular cloud cores using SPH]{The thermodynamics of collapsing molecular cloud cores using smoothed particle hydrodynamics with radiative transfer}

\author[S. C. Whitehouse \& M. R. Bate]{Stuart C. Whitehouse and Matthew R.
Bate\thanks{E-mail: scw@astro.ex.ac.uk, mbate@astro.ex.ac.uk}\\  School of Physics,  
University of Exeter, Stocker Road, Exeter
EX4 4QL}

\date{Accepted for publication in MNRAS}

\begin{document}
\maketitle

\begin{abstract}
We present the results of a series of calculations studying the collapse of
molecular cloud cores performed using a three-dimensional smoothed particle hydrodynamics code
with radiative transfer in the flux-limited diffusion approximation.   The
opacities and specific heat capacities are identical for each calculation. 
However, we find that the temperature evolution during the simulations varies
significantly when starting from different initial conditions. Even
spherically-symmetric clouds with different initial densities show markedly
different development. We conclude that simple barotropic equations of state
like those used in some previous calculations provide at best a crude
approximation to the thermal behaviour of the gas.  Radiative transfer is
necessary to obtain accurate temperatures.
\end{abstract}


\section{Introduction}

Radiative transfer is an important phenomena in star formation.
Radiation sets the temperature of the gas during the collapse of a molecular cloud core.
This both influences the degree of fragmentation of the cloud, and sets the minimum mass of 
brown dwarfs \citep[the opacity limit for fragmentation;][]{LL1976}.
Once protostars have formed in a cloud, radiative and mechanical feedback from them can affect subsequent star formation.  Such feedback mechanisms include protostellar jets and outflows 
from low-mass stars, and ionisation from massive stars which creates HII regions and destroys a 
cloud.

Computer simulations are vital in our efforts to understand the complex problem of star formation.
Many previous simulations have used the smoothed particle hydrodynamics method
(SPH) \citep[e.g.][]{GM1981,PCDNDW1991MmSAI,BMBAB1991,NP1993,BBP1995,KBB1998}.
Other methods used are typically based around grid-based codes  \citep[e.g.][]{L1969b,BB1979,BM1992,BB1993,TKMHHG1997,BFKM2000}.
SPH is a Lagrangian method first developed by
\citet{L1977} and \citet{GM1977} \citep[see][for a review]{M1992}. It
approximates the fluid as a series of discrete fluid elements denoted by
individual SPH particles and uses interpolation to obtain the fluid variables at
any point in the simulation. SPH is conceptually simple to understand, and can
naturally adapt its resolution to the local density distribution, unlike
grid-based codes which require complex adaptive-mesh refinement algorithms to
perform the same task. This property makes it ideal for use in star formation,
where densities may range over many orders of magnitude in a single simulation. 

Despite these advantages, few attempts have been made to include radiative transfer 
into SPH \citep{L1977,B1985,B1986,OW2003,WB2004}, and until recently 
\citep*{BCV2004,BCV2005} SPH with radiative transfer has not been applied to star
formation. Instead, many past simulations have simply used isothermal or
barotropic equations of state to model the collapse of a molecular cloud.  
The former is only valid up to densities of
$\sim 10^{-13}$ g~cm$^{-3}$ at which point the cloud traps radiation efficiently
enough for the cloud to begin to heat up.  The latter is usually based on the
evolution of the temperature at the highest density during the collapse of
spherically symmetric clouds as calculated using radiative transfer 
\citep[e.g.][]{L1969b,WN1980,MI2000}.
However, a barotropic equation of state can at
best only hope to provide an adequate description of temperature at the density
maximum; it is unlikely to give an accurate temperature distribution during a
three-dimensional calculation with complex density and velocity structure.
Indeed, \citet{BFKM2000} performed grid-based calculations of the collapse of a 
molecular cloud core both with a barotropic equation of state and with 
radiative transfer in the Eddington approximation and found they differed somewhat.
However, they did not examine in detail how the relation between temperature
and density differed from that of the barotropic equation of state spatially 
and temporally, or its dependence on initial conditions.

\citet*{WBM2005} recently presented an implicit algorithm for calculating
radiative transfer using the flux-limited diffusion approximation within the SPH
formalism.  This paper describes a three-dimensional implementation of this
algorithm and uses it to examine the thermodynamics during the collapse of 
molecular cloud cores. Section \ref{sec:method}
describes the changes necessary to the radiative transfer algorithm of
\citet{WBM2005} for use in three dimensions and the initial conditions for our
star formation calculations.  Section \ref{sec:results} presents the results of
simulations of the collapse of molecular cloud cores with different initial
conditions and examines the evolution of their temperature structure. Finally,
section \ref{sec:conclusions} summarises the main conclusions of this paper.

\section{Method and initial conditions}
\label{sec:method}

The code used in this paper is based on that of Bate \citep{B1995,BBP1995}, which originated
from that of Benz \citep{B1990c,BCPB1990}.  The code includes individual timesteps, and 
uses a tree to calculate self-gravity. The smoothing lengths of particles are variable in 
time and space, subject to the constraint that the number 
of neighbours for each particle must remain approximately 
constant at $N_{\rm neigh}=50$. We use the standard form of artificial viscosity 
(Monaghan \& Gingold 1983; Monaghan 1992) with strength 
parameters $\alpha_{\rm_v}=1$ and $\beta_{\rm v}=2$.
To perform radiation transport, the
algorithm from \citet{WBM2005}, adapted for use in three
dimensions, was added. 
The code has been parallelised by M.\ Bate using OpenMP.

\subsection{Three-dimensional flux-limited diffusion}

In a frame co-moving with the fluid, and assuming local thermal equilibrium
(LTE), the equations governing the time-evolution of radiation hydrodynamics
(RHD), integrated over frequency, are
\begin{equation}
\label{rhd1}
\frac{D\rho}{Dt} + \rho\mbox{\boldmath $\nabla\cdot v$} = 0~,
\end{equation}
\begin{equation}
\label{rhd2}
\rho \frac{D\mbox{\boldmath $v$}}{Dt} = -\nabla p + \frac{\mbox{$\kappa\rho$}}{c} \mbox{\boldmath $F$}~,
\end{equation}
\begin{equation}
\label{rhd3}
\rho \frac{D}{Dt}\left( \frac{E}{\rho}\right) = -\mbox{\boldmath $\nabla\cdot F$} - \mbox{\boldmath $\nabla v${\bf :P}} + 4\pi \kappa \rho B - c \kappa \rho E~,
\end{equation}
\begin{equation}
\label{rhd4}
\rho \frac{D}{Dt}\left( \frac{e}{\rho}\right) = -p \mbox{\boldmath $\nabla\cdot v$} - 4\pi \kappa \rho B + c \kappa \rho E~,
\end{equation}
\begin{equation}
\label{rhd5}
\frac{\rho}{c^2} \frac{D}{Dt}\left( \frac{\mbox{\boldmath $F$}}{\rho}\right) = -\mbox{\boldmath $\nabla\cdot${\bf P}} - \frac{\mbox{$\kappa \rho $}}{c} \mbox{\boldmath $F$}~
\end{equation}
\citep{MM1984,TS2001}. In these equations, 
$D/Dt \equiv \partial/\partial t + \mbox{\boldmath $v \cdot \nabla$}$  is 
the convective derivative.  The symbols $\rho$, $e$, {\boldmath $v$} 
and $p$ represent the material mass density, energy density, 
velocity, and scalar isotropic pressure respectively.  The total 
frequency-integrated radiation energy density, momentum density (flux)
and pressure tensor are represented by $E$, {\boldmath $F$}, and {\bf P},
respectively.  In this paper, we use a grey opacity, $\kappa$ (i.e., it is independent of frequency).

To solve radiation transport within SPH, we evolve both the specific internal
energy of the gas $u$, and the specific radiation energy $\xi=E/\rho$ implicitly
using the algorithm of \citet{WBM2005}. The principle difference between the
one-dimensional and three-dimensional forms of the radiative transfer equations
is the radiation pressure term (the second term on the right-hand side of
equation \ref{rhd3}). In one-dimension this term can be written using the
divergence of the  gas velocity.  However, in three dimensions this becomes the
tensor product  $\mbox{\boldmath $\nabla v${\bf :P}}$, where the Eddington pressure
tensor {\bf P} is given by 
\begin{equation}
\mbox{\bf P} = \mbox{\boldmath $f$} E.
\end{equation}
The components of the Eddington
tensor {\boldmath $f$}  are  given by 
\begin{equation}
\mbox{\boldmath $f$} = \frac{1}{2} \left( 1 - f \right) 
\mbox{\boldmath $I$} + \frac{1}{2} \left( 3f - 1 \right) 
\mbox{\boldmath $\hat{n}$} \mbox{\boldmath $\hat{n}$}.
\end{equation}
Here $f$ is the dimensionless scalar Eddington factor, and
\mbox{\boldmath $\hat{n}$} is the unit vector in the direction of the
energy density gradient $\nabla E / | \nabla E |$. 
For example, in Cartesian coordinates, the
first component of the tensor is given by
\begin{equation}
\frac{1}{2} (1-f) + \frac{\frac{1}{2} (3 f -1)
\left( \frac{\partial E}{\partial x} \right)^2 }{| \nabla E |^2 } ~ \xi,
\end{equation}
and the next by
\begin{equation}
\frac{\frac{1}{2} (3 f -1) \left( \frac{\partial E}{\partial x} \right) \left(
\frac{\partial E}{\partial y} \right) }{| \nabla E |^2 } ~ \xi,
\end{equation}
and so on.

The scalar Eddington factor $f$ is related to the flux-limiter $\lambda$, by the expression
\begin{equation}
\label{fld5}
f = \lambda + \lambda^2 R^2,
\end{equation}
where $R = |\nabla E|/(\kappa\rho E)$. In this paper, we choose
the flux limiter of \citet{LP1981}
\begin{equation}
\lambda(R) = \frac{2+R}{6 + 3R + R^2}.
\end{equation}
In the optically thick limit, $R\rightarrow 0$, $\lambda \rightarrow 1/3$, and {\bf P} becomes isotropic.

\subsection{Specific heat capacity and opacity}
 
The above equations are closed by the application of an equation of state for
the gas. We use the ideal gas equation of state
\begin{equation}
\label{eqn:eos}
p = \frac{R_{\rm g}}{\mu} \rho T_{\rm g},
\end{equation}
where $R_{\rm g}$ is the gas constant, $\mu$ is the mean molecular mass and the gas temperature is $T_{\rm g} = u / c_{\rm v}$.  The temperature of the radiation is given by $T_{\rm r}=(E/a)^{1/4}$.

We use the specific heat capacity $c_{\rm v}$ of \citet{BB1975}, 
which accounts for the dissociation of molecular hydrogen, and the ionization
of both hydrogen and helium. It omits any contribution due to metals. 
\begin{eqnarray}
\label{eqn:cv}
c_{\rm v} =&\displaystyle X \left( 1 - y \right) E(H_2) + \left[
1.5 X \left( 1 + x \right) y + 0.375 Y \left( 1 + z_{\rm 1} + z_{\rm 1} z_{\rm
2} \right) \right] R_{\rm g}  \nonumber
\\&+ X \left( 1.304 \times 10^{13} x + 2.143 \times
10^{12} \right) y/T_{\rm g} + \nonumber \\ & \displaystyle Y \left( 5.888 \times
10^{12} \left( 1 - z_{\rm 2} \right) +  1.892 \times 10^{13} z_{\rm 2} \right)
z_{\rm 1}/T_{\rm g},
\end{eqnarray}
where $X$ and $Y$ are the mass fractions of hydrogen and helium respectively
(in the simulations presented in this paper, $X=0.70$ and $Y=0.28$), $y$ is
the dissociation fraction of hydrogen, $x$ the ionisation fraction of hydrogen,
and $z_{\rm 1}$ and $z_{\rm 2}$ are the degrees of single and double ionisation
of helium, respectively. $E(H_{\rm 2})$ gives the contribution to the
specific heat capacity from molecular hydrogen \citep{BB1975}. The ionisation
fractions are calculated using the Saha equation. 

The variation in mean molecular mass $\mu$ with temperature was also taken from
\citet{BB1975} to be
\begin{equation}
\label{eqn:mu}
\mu^{-1} = \left[ 2X \left( 1 + y + 2 x y \right) + Y \left( 1 + z_{\rm 1}
 + z_{\rm 1} z_{\rm 2} \right) \right] /4.
\end{equation}

The opacity table from \citet{A1975} (the fourth King model) was used for
the gas opacity and \citet*{PMC1985} for the dust opacity. The opacity  values
were stored in a table containing the opacity at various  temperatures and
densities up to 10,000 K and 1 g cm$^{-3}$ respectively, and bilinear
interpolation in log-space was used to get the required value. Above 10,000 K,
the opacity was taken to be the lesser of extrapolating from the
last two points in the table, or Kramer's opacity
\begin{equation}
\label{eqn:kramer}
\kappa_{\rm K} = 1.2512 \times 10^{22} \rho T^{- \frac{7}{2}},
\end{equation}
until the electron scattering opacity $\kappa_{\rm es} = 0.4~$cm$^2$ g$^{-1}$
becomes the dominating opacity source at very high temperatures.

For temperatures above a few million K, the opacity, 
specific heat capacity and mean molecular mass become constant because the
contribution of metals is neglected. 
These three quantities were updated during every iteration of the implicit scheme. 

\subsection{Initial conditions}

The simulations described in this paper used three different sets of initial
conditions.  The first was the spherically symmetric collapse of \citet{BM1992}.
They began with a $1.2~{\rm M}_{\odot}$ sphere of uniform density
$\rho = 1.7 \times 10^{-19}$ g cm$^{-3}$, giving a free-fall time of $t_{\rm ff}=1.61\times 10^{5}$ years. The cloud radius was $1.5 \times
10^{17}$ cm, and the initial temperature of both the gas and the radiation
field was $10$ K. 

The other two initial conditions were based on those of \citet{BB1979}.  
They began with a $1.0~{\rm M}_{\odot}$ sphere of density $\rho = 1.44 \times 10^{-17}$ g cm$^{-3}$ 
and radius  $3.2 \times 10^{16}$ cm, giving a free-fall time of $t_{\rm ff}=1.75\times 10^4$ years. Their cloud was initially in solid-body
rotation with an angular velocity of $1.6 \times 10^{-12}$ rad s$^{-1}$.  Superimposed
on the underlying density was an $m=2$ density perturbation satisfying
\begin{equation}
\rho = \rho_0 [1 + 0.5\cos(m\phi)]
\end{equation}
where $\phi$ is the angle in the plane perpendicular to the axis of rotation.
The ratio of thermal energy to magnitude of the gravitational energy was initially
0.26 and the ratio of rotational energy to the magnitude of the gravitational energy
was 0.20.  Using our equation of state, the initial temperature of the gas and radiation
were therefore set to 12 K.  We performed calculations with initial conditions identical 
to those of \citet{BB1979},
and also a spherically symmetric version in which there was no $m=2$ density perturbation
applied and the cloud was not rotating.

We performed simulations with 5000, 50,000, 150,000 and 500,000 particles. Of
these, simulations with 50,000 particles and greater resolve the Jeans mass
according to the criteria set out in \citet{BB1997}.   The calculations were performed
on the United Kingdom Astrophysical Fluids Facility (UKAFF). The highest resolution 
(500,000 particle) calculation took a total of approximately 3000 CPU hours (running across multiple processors).

\section{Results}
\label{sec:results}

\begin{figure}
\centerline{\psfig{figure=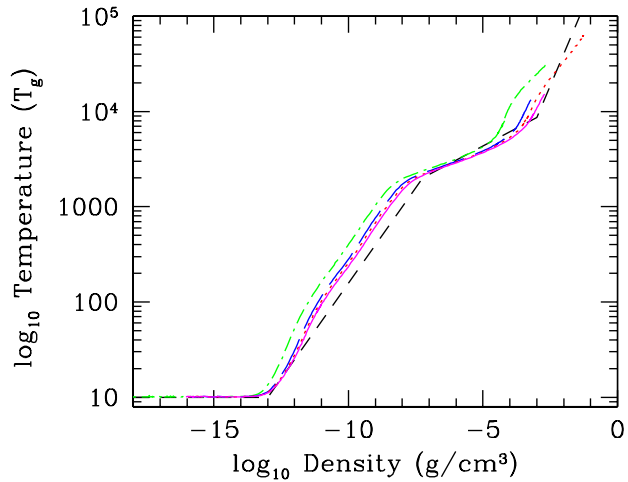,width=8.5truecm}}
\caption{\label{fig:BMres}  We show the evolution of the maximum temperature and
density during the calculations that start from the initial conditions of
\citet{BM1992} for different resolutions.  The calculations were performed using
5000  (dot-dashed green line), 50,000 (long-dashed blue line), 150,000 (dotted red line) and 500,000 (solid magenta line) particles. 
The results converge as the resolution is increased, with 50,000 or greater
numbers of particles required to achieve reasonable results.   For comparison,
the short-dashed line gives the temperature-density relation of the barotropic
equation of state used by \citet{B1998}. }
\end{figure}

A brief summary of the expected behaviour during a spherically symmetric
collapse is as follows \citep[see][for more details]{BS1968,L1969b}. A Jeans-unstable
uniform cloud of molecular gas will begin to collapse under its own self-gravity. 
The collapse occurs such that the density in the outer parts of the cloud falls off with radius $r$
as $\rho \propto r^{-2}$, while the uniform-density inner part of the cloud collapses to higher and
higher densities but contains a decreasing fraction of the mass as the collapse proceeds.
Initially, the collapse is isothermal, as the material is
optically thin. However, once the central region reaches a critical density ($\sim
10^{-13}$ g cm$^{-3}$) it becomes optically thick and starts to heat as energy
can no longer be radiated away. The central temperature and density both rise
rapidly, until thermal pressure can counter the collapse. This results in the
formation of a pressure-supported core in the centre of the cloud. Mass continues to infall onto this
core, increasing its mass and temperature.  While at low temperatures, the molecular hydrogen
behaves like a monatomic gas (the ratio of specific heats $\gamma=5/3$).  However, when the
temperature approaches $\sim 100$ K (at densities $\sim 10^{-11}$ g cm$^{-3}$)
additional degrees of freedom become available and $\gamma=7/5$. 
This phenomenon which is included
in our specific heat capacity is not typically included in the simple barotropic equation
of state. Once the core reaches $\sim 2000 $ K, the molecular hydrogen
begins to dissociate. As energy is diverted into this dissociation rather than
thermal support, the core begins a second collapse phase. Once the hydrogen has
been fully dissociated, it forms a second (``stellar'') core, once again
supported by thermal pressure. This core continues to increase mass and
temperature as material falls onto it from the envelope.

\begin{figure}
\centerline{\psfig{figure=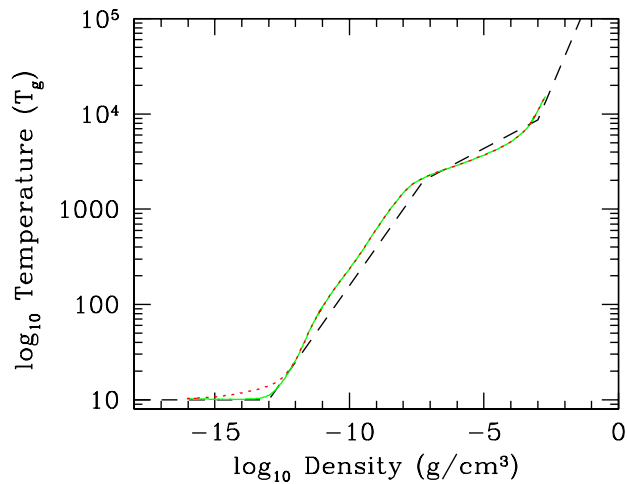,width=8.5truecm}}
\caption{\label{fig:BM500k} We show the evolution of the maximum gas (solid green
line) and radiation (dotted red line) temperatures at the maximum density during the
\citet{BM1992} calculation.  This was our highest resolution (500,000 particles)
calculation. The gas and radiation temperatures are identical except during the
transition from optically thin to optically thick.    For comparison, the short-dashed
line gives the temperature-density relation of the barotropic equation of state
used by \citet{B1998}.}
\end{figure}

\begin{figure*}
\centerline{\psfig{figure=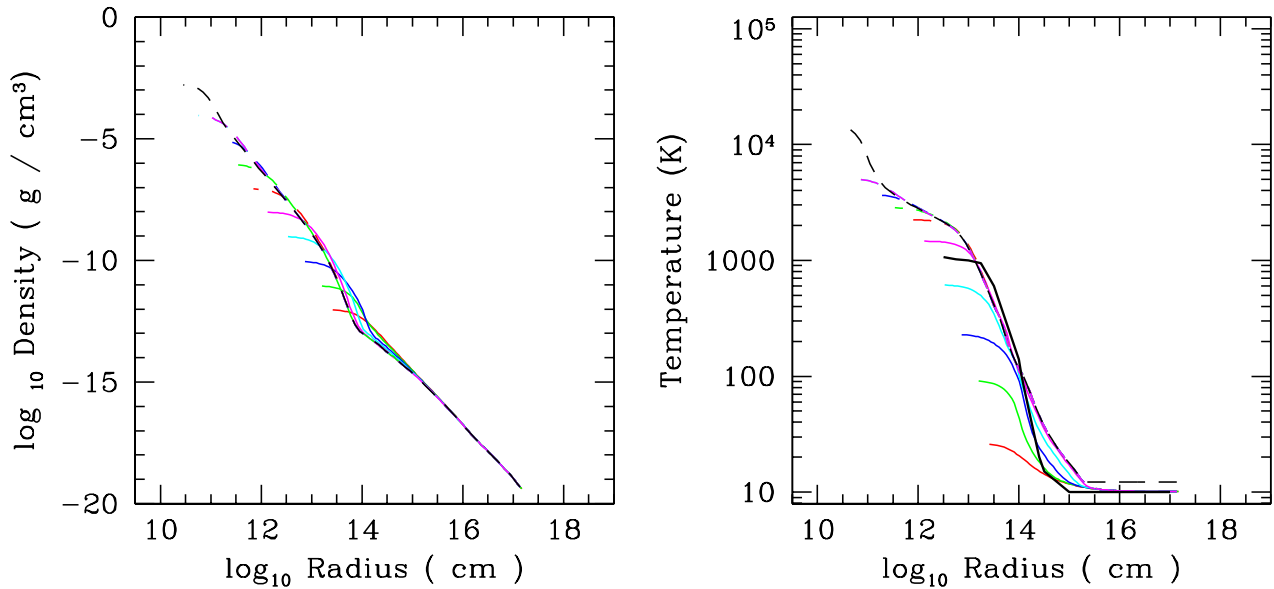,width=18.0truecm}}
\caption{Radial profiles of density (left panel) and gas temperature (right
panel) at various stages during the highest resolution (500,000 particles)
\citet{BM1992} calculation.  A profile is plotted every order of magnitude in
maximum density from $10^{-12}$ to $10^{-3}$ g~cm$^{-3}$ and finally at $2
\times 10^{-3}$ g~cm$^{-3}$.   For comparison, the thick solid black lines give
the profiles from \citet{BM1992} when the temperature in their calculation was
$\approx 1000$ K. \label{fig:BMmix}}
\end{figure*}

\begin{figure}
\centerline{\psfig{figure=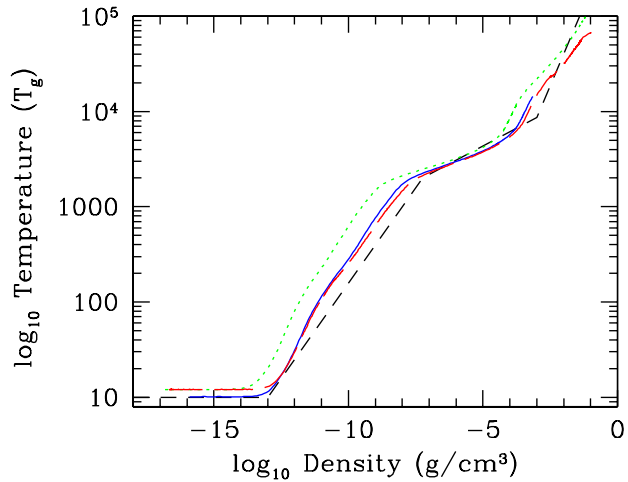,width=8.5truecm}}
\caption{\label{fig:50k} The evolution the maximum temperature and density
during calculations starting from three different sets of initial conditions. 
The three calculations are the spherically-symmetric (dotted green line) and $m=2$
rotating (long-dashed red line)  \citet{BB1979} collapses and the spherically-symmetric
\citet{BM1992} collapse (solid blue line).    All calculations were performed using
50,000 particles.  Note that different initial conditions can lead to very
different temperature evolution.  For comparison, the short-dashed line gives the
temperature-density relation of the barotropic equation of state used by
\citet{B1998}.}
\end{figure}

\begin{figure}
\centerline{\psfig{figure=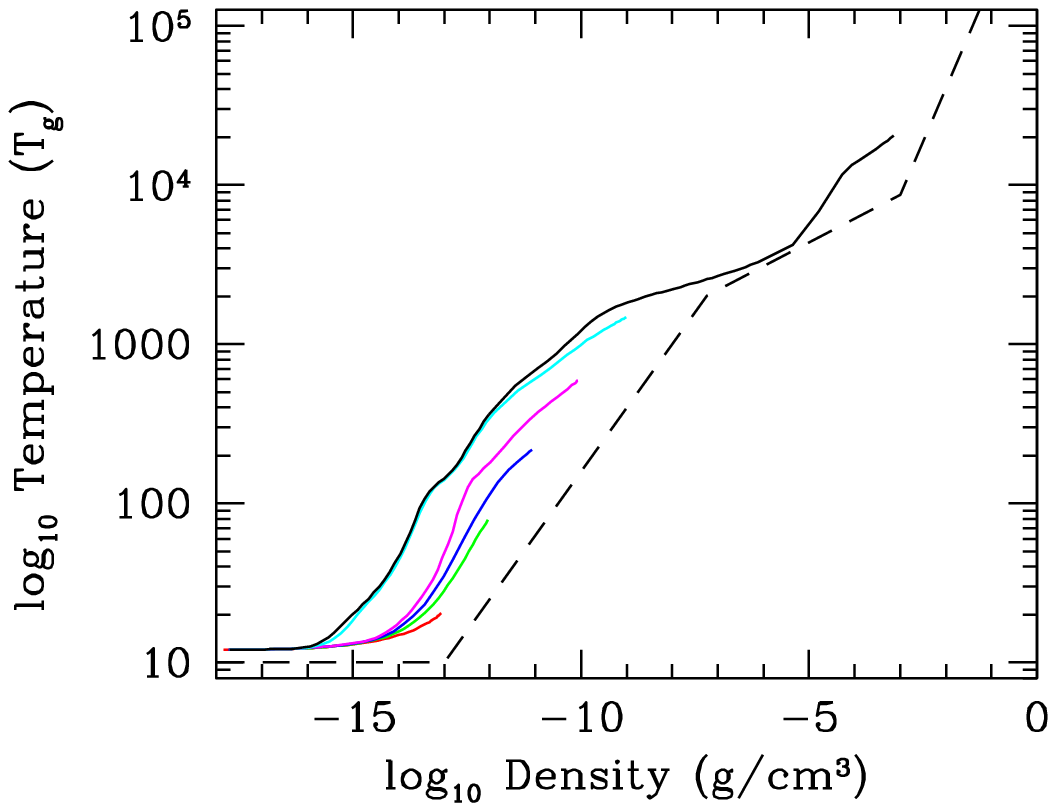,width=8.5truecm}}
\caption{ \label{fig:BBsnap}  Snapshots of temperature versus density at various
stages during the 50,000 particle spherically-symmetric \citet{BB1979}
calculation.  A snapshot is plotted every order of magnitude in maximum density
from $10^{-13}$ to $10^{-9}$ g~cm$^{-3}$ and finally at $10^{-3}$ g~cm$^{-3}$. 
As with the \citet{BM1992} initial conditions, the temperature differs by more
than an order of magnitude from that given by the barotropic equation of state
(dashed line) at late times.} 
\end{figure}

\begin{figure} 
\centerline{\psfig{figure=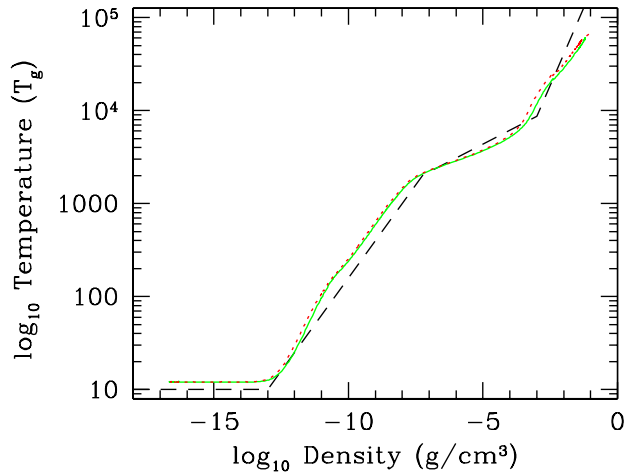,width=8.5truecm}}
\caption{\label{fig:BBbinarycomp} The evolution of the maximum temperature and
density during the calculations that start from the initial conditions of
\citet{BB1979}, including the $m=2$ density perturbation and rotation, for two
different resolutions.  The calculations were performed using 50,000 (dotted red line)
and 150,000 (solid green line) particles.  The two resolutions give similar results. 
For comparison, the dashed line gives the temperature-density relation of the
barotropic equation of state used by \citet{B1998}.} 
\end{figure}

We present results showing the evolution with time of our calculations in Figures
\ref{fig:BMres} to \ref{fig:BBbin150k}. The short-dashed black line in these figures is the
barotropic equation of state used by \citet{B1998}, given for comparison.  Unless otherwise
stated, we plot the temperature of the gas rather than that of the radiation field.

As far as we are aware, these are the first three-dimensional radiative transfer
calculations to follow the collapse of a molecular cloud core beyond the
formation of the first pressure-supported core and the dissociation of molecular
hydrogen to the formation of the stellar core.  \citet{Boss1984} performed one-
and two-dimensional radiative transfer calculations that followed the collapse
of a molecular cloud core well beyond the formation of the first
pressure-supported core.  However, in order to accomplish this he had to employ
several numerical artifices, begging the question of how we are able to perform
{\it three-dimensional} calculations.  Boss' first artifice was to implement two
regions of his grid that were evolved using different timesteps in order to
follow the first core over many dynamical times using short timesteps but
simultaeously model the envelope with larger timesteps.  He also damped out
oscillations of the first core to decrease the amount of computational effort
required.  In our calculations, we encountered no difficulties in modelling the
cloud collapse through the first core phase and onto the formation of the
stellar core.  We believe there are three reasons for this.  First, our code
employs individual timesteps \citep{B1995,BBP1995} for each particle (i.e.
similar to, but even more efficient than, the way Boss evolved his calculation
in with two spatially distinct timesteps).  By the end of our highest resolution
calculation, some particles within the stellar core were being evolved using
timesteps of less than 1/40 of a second ($\approx 10^5$ times smaller than those
in the outer parts of the cloud).  Second, our calculations used the standard
form of SPH viscosity with none of the possible viscosity-reducing
formulations.  In each calculation, the first core undergoes oscillations after
its formation, as observed in earlier work, but the viscosity likely damped
these oscillations in a similar way to Boss' artifice.  Finally, in the twenty
years between Boss' and our calculations, computers have become very much
quicker.

\subsection{Results using Boss and Myhill initial conditions}

Figure \ref{fig:BMres} shows the evolution of maximum temperature and density 
during the \citet{BM1992} collapse calculations.  We performed the collapse with
four different resolutions.  With 5000 particles (green line) the cloud heats at an 
earlier stage than in the higher resolution simulations, presumably due to insufficient resolution. 
The 50,000 particle collapse (blue) is much cooler for a given maximum density,
while the two highest resolution simulations appear to be converging towards a single
curve.  We conclude that 50,000 to 150,000 particles (i.e. approximately the same number
as required to resolve the Jeans mass) are sufficient to model the
thermal behaviour reasonably accurately.  The evolution of maximum temperature and
density follows the barotropic equation of state (dashed line) in a qualitative sense, but 
there are still differences in temperature of up to $\approx 50$ percent between the
radiative transfer and the barotropic equation of state at various stages during the
collapse.

Figure \ref{fig:BM500k} shows the evolution of both the gas
temperature (green) and the radiation temperature
(red) for the highest resolution (500,000 particles) \citet{BM1992} collapse. The
radiation temperature is equal to the gas temperature at most stages, except
in the vicinity of the transition between the optically thin and thick regimes.  This
can be seen in the figure from densities of $\sim 10^{-15}$ to $10^{-12}$ g~cm$^{-3}$
when the cloud initially begins to heat.  Here the radiation temperature is
up to a factor of two greater than that of the gas.  Later in the calculation, gas that is 
optically thick has the same temperature as the radiation and the two remained 
coupled as the gas collapses to very high densities.  However, there is always a
region where the gas transitions from optically thin to optically thick where the temperatures
differ.

\begin{figure}
\centerline{\psfig{figure=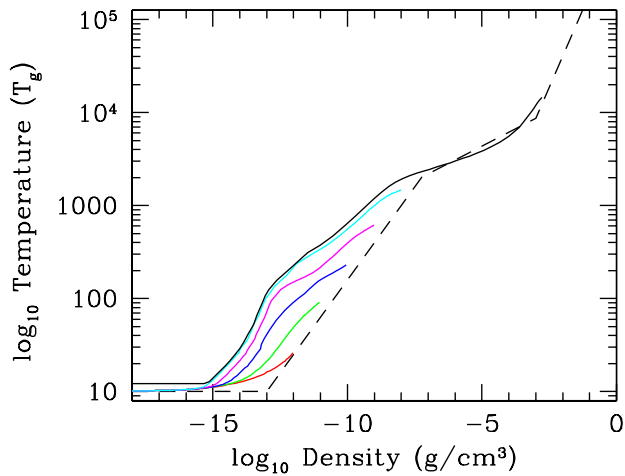,width=8.5truecm}}
\caption{\label{fig:BMsnap} Snapshots of temperature versus density at various
stages during the highest resolution (500,000 particles) \citet{BM1992}
collapse.  A snapshot is plotted every order of magnitude in maximum density
from $10^{-12}$ to $10^{-8}$ g~cm$^{-3}$ and finally at $2 \times 10^{-3}$ g~cm$^{-3}$. 
Note that at late times the temperature differs by more than an order of
magnitude from that given by the barotropic equation of state (dashed line).}
\end{figure}

Figure \ref{fig:BMmix} shows density (left) and gas temperature (right) plotted
against radius as the collapse passes various values of the maximum density.  
The first core can be seen at a radius of $\sim 10^{14}$ cm where the density and temperature 
gradients change abruptly.  Similarly, the transition to the stellar core can be seen at a radius
of $\sim 10^{12}$ cm.  \citet{BM1992} give the density and temperature profiles when the
central temperature reaches $\approx 1000$ K.  We reproduce their profiles as the thick solid black
lines in Figure \ref{fig:BMmix}.  Our results are in good agreement with theirs, the main difference being that our results give slightly higher temperatures at radii $\sim 10^{15}$ cm, presumably due to the fact that we use flux-limited diffusion while \citet{BM1992} use the Eddington approximation (the former retards radiation transport near the surface of the first core).

In Figure \ref{fig:BMsnap}, we plot temperature versus density at a series of snapshots
during the highest resolution calculation (since both temperature and density are monotonically 
decreasing functions of radius, we can plot these as lines for each snapshot).
The figure clearly shows that even if it were possible to construct a barotropic equation
of state to follow the evolution of maximum density and temperature, it would generally
underestimate the temperature of the gas away from the maximum.  In the figure, the
barotropic equation underestimates the temperature by more than an order of magnitude
at densities $\sim 10^{-13}$ g~cm$^{-3}$ at late times.

\subsection{Results using spherically symmetric Boss and Bodenheimer initial conditions}

Figure \ref{fig:50k} shows the evolution of the maximum value of temperature
and density for the three types of initial conditions, each performed with 50,000
particles. The spherically-symmetric \citet{BB1979} collapse (green line) begins
to heat at a much lower density than the \citet{BM1992} collapse (blue).  Essentially
the only difference between these calculations is the initial density of the two clouds,
the former is initially almost two orders of magnitude denser,
and yet the temperature evolutions are quite different.  Figure \ref{fig:BBsnap} shows
that, as with the \citet{BM1992} collapse, the spatial and temporal evolution of the temperature is
also complex.  Again, these results demonstrate that a barotropic equation of state cannot accurately
describe the temperature distribution during such cloud collapses.

\subsection{Binary star formation}

The result of the \citet{BB1979} initial conditions, including the $m=2$ density perturbation and rotation, is to form a wide binary with a separation of $\approx 1000$ AU \citep{BBP1995}.  Using these initial conditions, the evolution of the maximum temperature versus density is much cooler than that from its spherically-symmetric counterpart (see Figure \ref{fig:50k}, red line).  In fact, coincidentally, it is actually quite close to the evolution obtained with the \citet{BM1992} initial conditions.  There are likely to be two reasons for the cooler temperatures at the same maximum density in the rotating collapse.  First, the collapse to form the stars is delayed by the rotational support (they form at $t=1.30$ initial cloud free-fall times instead of just over one free-fall time for the spherically symmetric case).  The slower collapse allows more time for energy to be radiated away from the object.  Second, the presence of rotation means that the distribution of material around each of the collapsed objects is no longer spherically symmetric.  In particular, there are low density cavities along the rotation axes and high densities in the discs.  Radiation can more easily escape through these cavities than in the spherically symmetric case.

We also tested the effect of resolution on the \citet{BB1979} binary star initial conditions, using 
50,000 and 150,000 particles (Figure \ref{fig:BBbinarycomp}, red and green lines respectively).  As with the \citet{BM1992} initial conditions, increasing the resolution above 50,000 particles has little effect on the temperature evolution showing that the calculations are essentially converged.

\begin{figure}
\centerline{\psfig{figure=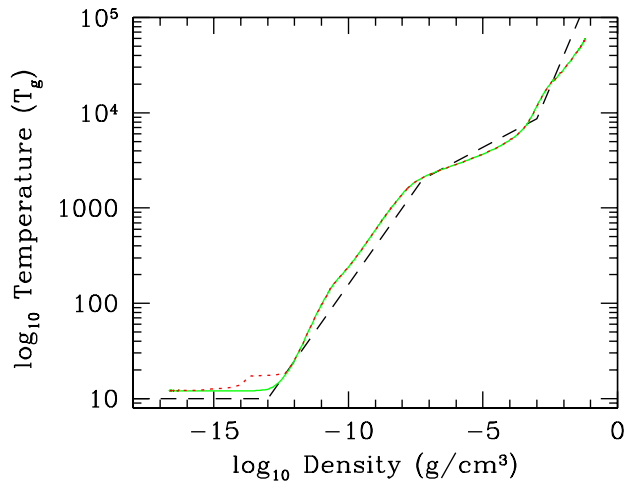,width=8.5truecm}}
\caption{\label{fig:BBbin150k} The evolution of the maximum gas (solid green line) and
radiation (dotted red line) temperatures at the maximum density during the
\citet{BB1979} calculation, including the initial $m=2$ density perturbation and
rotation.  The calculation was performed using 150,000 particles. The gas and
radiation temperatures are identical except during the transition from optically
thin to optically thick.    For comparison, the dashed line gives the
temperature-density relation of the barotropic equation of state used by
\citet{B1998}.}
\end{figure}

The difference in evolution between the radiation and the gas temperatures is
shown in Figure \ref{fig:BBbin150k}, in a manner similar to Figure
\ref{fig:BM500k}. Again the only significant difference occurs during the transition from the optically thin to optically thick regimes at densities around $10^{-15}$ to $10^{-12}$ g~cm$^{-3}$.

\begin{figure*}
\centerline{\psfig{figure=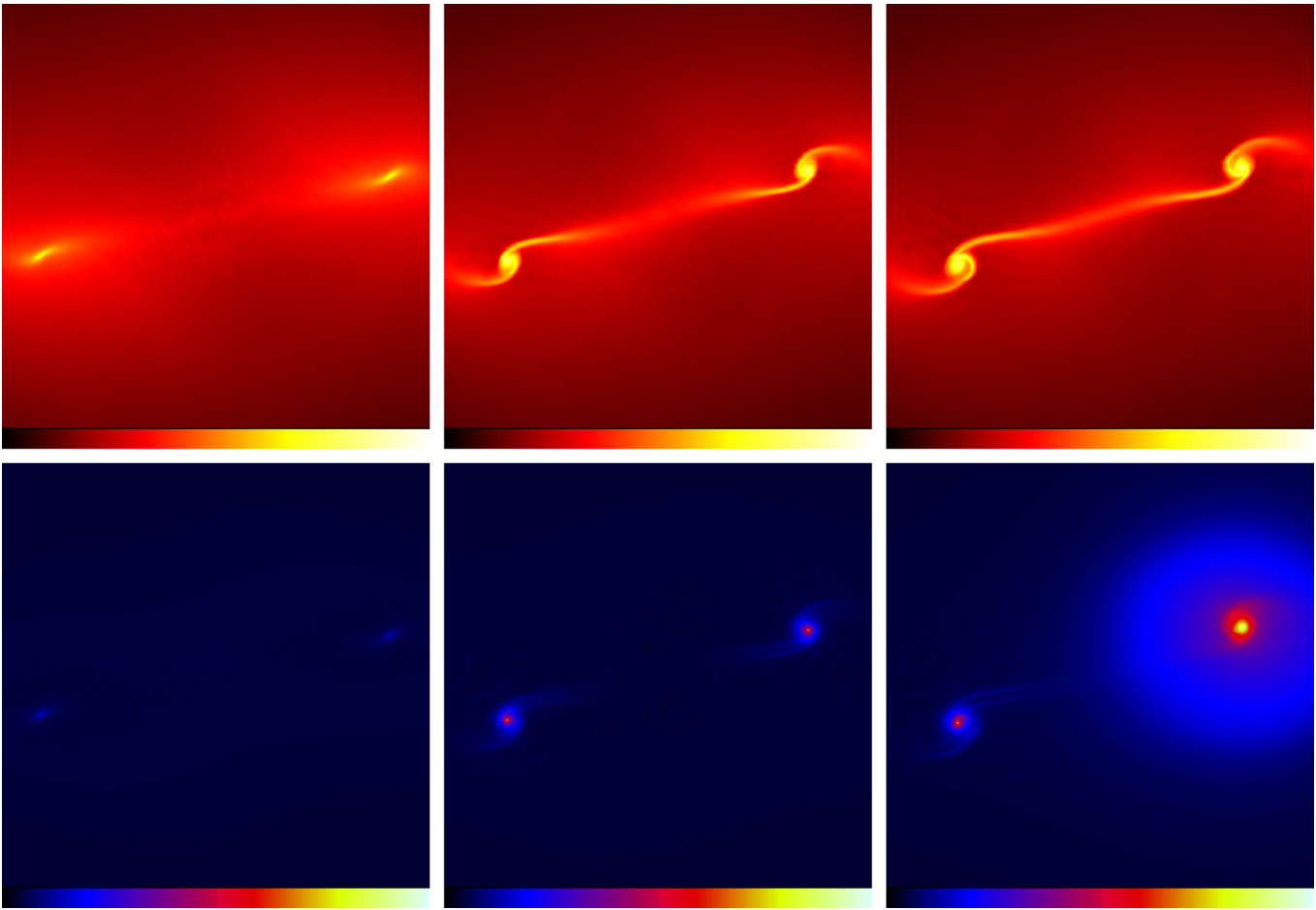,width=18.0truecm}}
\caption{\label{fig:BBbin150kcolour} Snapshots of the logarithmic column
density, $N$, (upper panels) and mass-weighted gas temperature, $T_{\rm g}$,
(lower panels) parallel to the rotation axis for the 150,000 particle
\citet{BB1979} calculation, including the initial $m=2$ density perturbation and
rotation.  The density scale covers $-1 < \log N < 3.5$ with N measured in
g~cm$^{-2}$ while the temperature scale ranges from $1 < \log T_{\rm g} < 3$.
The snapshots are given when the maximum densities are $10^{-12}$, $10^{-10}$,
and $10^{-1}$ g~cm$^{-3}$.  The corresponding times are $1.18$, $1.24$ and
$1.30$ initial cloud free-fall times. The molecular cloud core collapses to form
a protostellar binary system.  Note that the dense gas within the spiral arms is
cooler than the surfaces of arms (e.g., lower middle panel and the left-hand
object in the lower right panel).  In the lower right panel, the protostar on
the right has formed a stellar core and thus heated the surrounding gas
significantly more than the protostar on the left.  At the time plottted, the
left protostar  has not quite undergone molecular hydrogen dissociation to form
a stellar core. The two objects form stellar cores at slightly different times
because of  small asymmetries in the initial conditions. }
\end{figure*}

\section{Conclusions}
\label{sec:conclusions}

The results of our study show that the initial conditions in a molecular cloud
core such as the density and  velocity configurations have a large effect on the
evolution of the temperature as the cloud collapses. The evolution of the
maximum temperature and density during the collapse cannot be approximated by a
single barotropic equation of state.  Furthermore, even if a satisfactory
description of this evolution was able to be formulated by a simple equation of
state, it would do a very poor job of setting the temperature away from the
location of the density maximum.  We find the barotropic equation of state used
in many previous simulations \citep[e.g.][]{B1998} may underestimate the
temperature by up to a factor of two to three for the density maximum, and by
more than an order of magnitude in other parts of the cloud.   In principle,
this may seriously affect the evolution of a molecular   cloud core, even
altering how it fragments since the Jeans mass scales with gas temperature as  
$T_{\rm g}^{3/2}$. Therefore, if possible, radiative transfer should be used.

\section*{Acknowledgments}

SCW acknowledges support from a PPARC postgraduate studentship. MRB is grateful
for the support of a Philip Leverhulme Prize. The computations reported here
were performed using the UK Astrophysical Fluids Facility (UKAFF).

\bibliographystyle{mn2e}
\bibliography{MF1128rv}

\end{document}